# Threshold conduction in amorphous phase change materials: effects of temperature

J C Martinez,[1] Ronald A Coutu Jr,[2] Turja Nandy,[2] R E Simpson[1]

[1]Singapore Univ. of Technology and Design, 8 Somapah Rd, Singapore 487372
[2]Dept. of Electrical & Computer Engineering, Marquette Univ., Milwaukee, WI. 53233 USA

We emphasize the role of temperature in explaining the *IV* ovonic threshold switching curve of amorphous phase change materials. The Poole-Frankel conduction model is supplemented by considering effects of temperature on the conductivity in amorphous materials and we find agreement with a wide variety of available data. This leads to a simple explanation of the snapback in threshold switching. We also argue that low frequency current noise in the amorphous state originates from trains of moving charge carriers derailing and restarting due to the different local structures within the amorphous material.

**Keywords**: ovonic threshold switching, Poole-Frankel model, amorphous chalcogenide alloys, phase change materials, tunneling, Newton's law of cooling, negative differential resistance

## 1. Introduction

Fifty-two years ago, Stanford Ovshinsky announced the discovery of electrical ovonic threshold switching in amorphous chalcogenide alloys [1]. Initially bypassed, the discovery now promises to transform data storage, computing in memory, hardware neural networks, as well as other switching technologies [2]. An important issue inherent to all amorphous materials concerns the nature of electrical transport [3, 4]. Efforts to understand this mechanism have been in earnest in the past 20 years. Conduction in phase change materials (PCM) has two disparate manifestations: when the material is amorphous, the resistance is very large; while, when the material is crystalline, the resistance is orders of magnitude smaller. The characteristics of the crystalline case can be explained in terms of those of a doped semiconductor with a small band gap and Fermi level lying close to the valence band [5]. When the material becomes amorphous, it hosts a large number of localized states which contribute to conduction only in the presence of large external electric fields [6, 7]. Aspects of the Poole-Frankel (PF) model aid the understanding of conduction in this amorphous phase by accounting for the subthreshold conduction which compares well with measured current-voltage characteristics [8]. While it is widely accepted that threshold switching has an electronic origin [9], questions about the fast, thermal dynamics of nanoscale phase change devices have opened fresh debate [10]. Moreover, the role of direct tunneling or thermally assisted tunneling from a single defect into the valence band has been seriously questioned [11]. Although temperature is not entirely ignored in these discussions, we argue that much more attention to it is really required. Herein, we show that the parametrizing of the current-voltage (*IV*) plots is best achieved by incorporating temperature effects. Moreover, we will see that the parametrization works well even when temperature does not play a dominant role. Further, we show how these effects appear in negative differential resistance.

Two effects are frequently associated with conduction in amorphous PCM: $1/f$ noise and the voltage snapback. Typically, $1/f$ noise in an electronic device accompanies the change in carrier trapping and de-trapping which manifests as temporal fluctuations. In the case of amorphous chalcogenide PCMs it has been attributed to bond length/angle induced fluctuations of the mean trap energy [12]. The voltage snapback has been explained in terms of the electron energy increase due to the electric field which enhances conduction. This energy increase coupled with the E-field non-uniformity in the amorphous region leads to dielectric breakdown [13]. For the voltage snapback we propose, in line with our discussion above, an explanation that gives greater importance to the role of temperature. Moreover, in the case of the snapback we focus on its occurrence at the very onset of conduction for amorphous PCMs, in which case phase change has *not* yet occurred, i.e. the PCM is structurally amorphous but conducting. As for $1/f$ noise we direct attention to *very low frequency* noise, below 100 Hz, which we link neither with single electrons nor holes, but with the short-range order of amorphous materials. From this perspective, it may have relevance to the old debate on Stark-Wannier ladders [14].

A word about the difference between crystalline and amorphous materials is helpful in framing our discussion. While periodicity, symmetry and Bloch's theorem are central to crystals, a dearth of ordered atomic structure in amorphous materials naturally shifts emphasis to short-range bonding interactions [6, 7]. That said, amorphous materials are only broadly disordered since they retain the same order as crystals when viewed locally but not globally. The amorphous state of germanium telluride, a representative PCM, crystallizes when its temperature is above 500 K [15].

## 2. Poole-Frankel picture

If, for a moment, we imagine a regular crystalline environment, a constant electric field $\mathcal{E}$ exerts a potential $-q\mathcal{E}x$ (figure 1(a)), which can be broken into its non-periodic and periodic components. Periodicity is preserved by the second component as it shares in the crystal's periodicity, while the first does not. In the absence of the field, however, an electron in the crystal sees the Coulomb potential of the positively charged ions of the crystal as a series of regularly spaced potential humps. For such an electron with wave number $k_0$ at $t = 0$ under the electric field $\mathcal{E}$, it can be shown that the wave function at time $t$ is given by [16]

$$\Psi(x,t) = e^{ik_0x}u_{n,k_0+q\mathcal{E}t/\hbar}(x)\exp\left(\frac{i}{q\mathcal{E}}\int_{k_0}^{k_0+q\mathcal{E}t/\hbar}\epsilon(k',\mathcal{E})\,dk'\right) \quad (1)$$

where $\epsilon\,(k,\mathcal{E})$ is the energy eigenvalue for fixed wave number $k$ and $u_{n,k}(x)$ are electric-field-dependent Bloch states ($n$ = band index) and the overall phase factor at the right comes from the non-periodic component. When the field is not very large, bands form which are not mixed by the electric field (this is the Stark-Wannier debate mentioned above).

In the amorphous case with its short-range order and hence loss of periodicity, the waves undergo destructive interference with information from the phase factor being lost, but the field dependent Bloch states largely remain intact in the vicinity of any given site [17]. Thus, despite the loss of phase information, short-range order in amorphous material under an electric field is retained by the field-dependent Bloch states. Close to any given site, the PF picture sees classical electrons being thermally emitted and hopping over the top of the Coulomb-barrier humps between charge centers (figure 1(b)) which have been concomitantly distorted by the external electric field (figure 1 (c)) [18, 19].

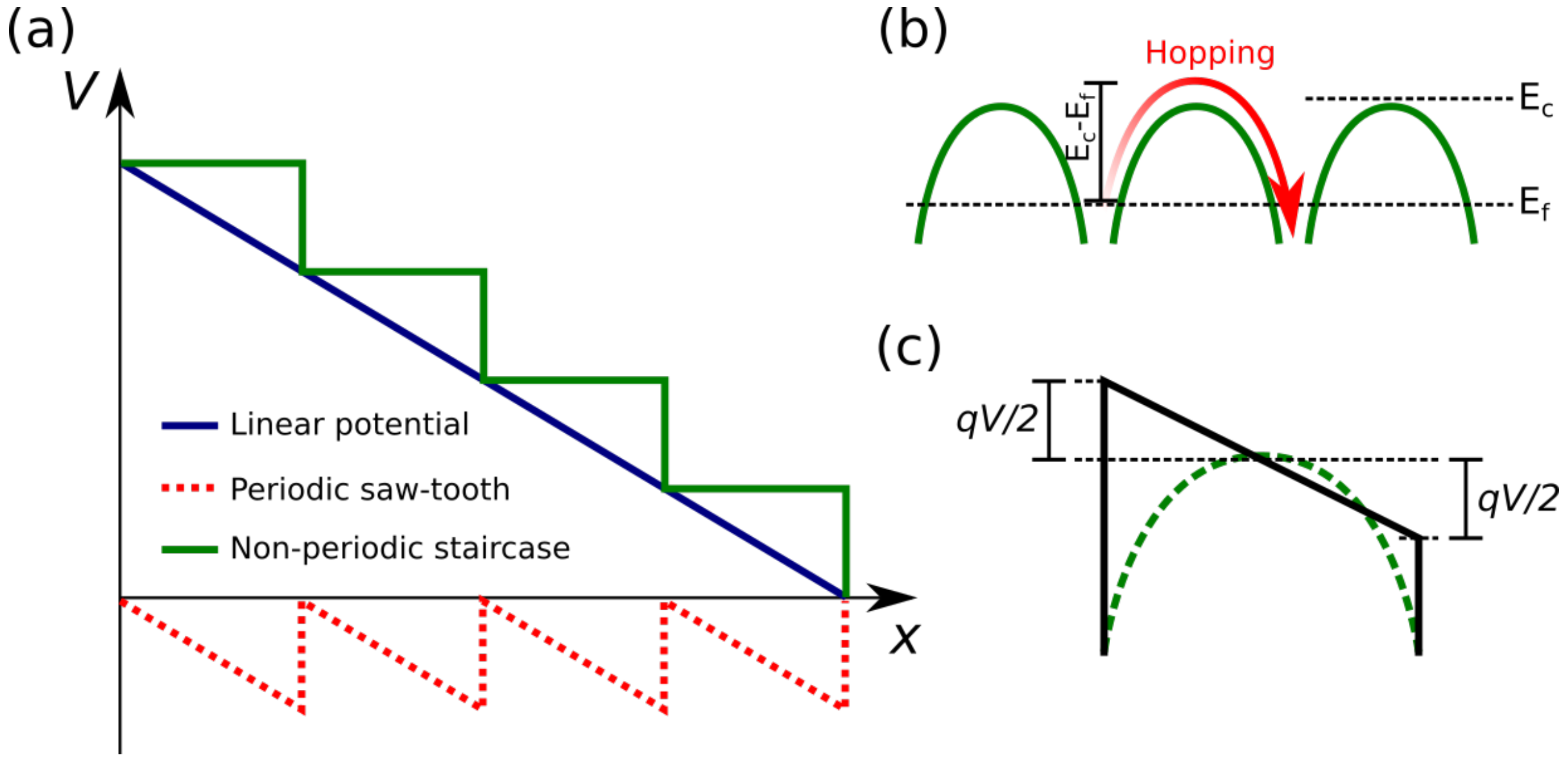

**Figure 1** (a) In a crystal, a linear external electric potential can be broken into its non-periodic staircase component and periodic saw-tooth component. (b) In red, the PF picture of electron hopping between traps originating from potential humps created by the Coulomb interaction between trapping centers. The barrier height $E_C - E_F$ is the potential difference between the conduction band and Fermi level; (c) Although long-range order is lost in amorphous materials, in a given neighborhood, the periodic component of the field together with the Coulomb interaction form a trapezoidal potential barrier with the left potential raised by $qV/2$ and the right potential lowered by the same amount ($V$ = applied voltage). The dashed curve is the Coulombic hump of (b). See text below.

Tunneling into the potential barrier is made possible through the electron thermal energy and the external field. During a measurement, the device temperature need not be constant, and this possibility implies a system that is not in equilibrium. For a given DC-current measurement we will assume steady-state conditions. Then despite the issue of non-equilibrium and the lack of long-range order in a-PCM, under steady-state conditions we can extrapolate local measurements to apply to typical device-sized regions.

## 3. Current density

We outline a derivation of the current density, $j$, to highlight important features in the conduction mechanism in a solid. Bardeen gave an expression for the electron current density $j$ for electrons in a state $a$ on one side of an interacting region transiting to a state $b$ on the opposite side

$$j = \frac{4\pi q}{\hbar} \sum_{k_t} \int |M_{ab}|^2 \rho_a \rho_b f_a (1 - f_b) dE \quad (2)$$

where $M_{ab}$ is the matrix element for the transition, $\rho_a$, $\rho_b$ are densities of states, with corresponding occupation probabilities $f_a, f_b$, and $q$ is the electron charge [20]. The integral over energy $E$ is taken at a fixed transverse wave number $k_t$. This result is given without explicitly assuming the Bloch theorem or the specific physical mechanism involved except that the wave functions decay exponential away from the interaction region. Moreover, spin and current in the reverse direction have been accounted.

At this point we may consider the tunneling mechanism. The matrix element takes on the form $|M_{ab}|^2 = (\frac{\hbar^2}{2m})^2 \frac{(k_z)_a}{L_a} \frac{(k_z)_b}{L_b} |T|^2$, where $(k_z)_{a,b}$ are wave numbers in the direction of the current (z-direction), $L_{a,b}$ are length scales for states $a$ and $b$, respectively, and $|T|^2$ is the usual Gamow tunnelling factor [21]. In the independent particle model, which assumes fairly sharp boundaries at the ends of the potential barrier, the expressions for the density of states cancel all the factors to the left of $|T|^2$ so equation (2) simplifies to $j = \frac{2q}{h} \sum_{k_t} \int |T|^2 (f_a - f_b) dE$. Although the independent particle model is not consistent with some experimental results, it is not expected to fail for normal metals and semiconductors[21]. Writing the occupation probabilities in the form $f_a = k_B T \frac{d}{dE_t} \ln(e^{\beta(E_F - E)} + 1)$, $E_F =$ Fermi energy, $\beta = (k_B T)^{-1}$, the sum over transverse momenta can be carried out using $\frac{1}{A} \sum_{k_t} \rightarrow \frac{d^2 k_t}{(2\pi)^2}$ to obtain the Esaki-Tsu formula [22]

$$j = -\frac{q m k_B T}{2\pi^2 \hbar^3} \int dE_l \, |T|^2 \ln\left(\frac{e^{\beta(E_F - E)} + 1}{e^{\beta(E_F - E - qV)} + 1}\right) \quad (3)$$

$m$ being the carrier mass. We had assumed that the initial and final states $a$ and $b$ differ in energy by the electric-field potential $qV$, $V =$ applied voltage.

Let us now turn to amorphous materials for which detailed balance and Boltzmann statistics are assumed. Following the trap-limited conduction model [8] we envisage conduction electrons separated by a potential barrier of width $d$ and height $E_C - E_F$, i.e. the difference between the conduction and Fermi energy (see figure 1(b)). The experimental set-up is generally referred to as a device. When the electric field pointing in the z-direction is turned on, the electrons see a trapezoidal potential across the barrier of the form $\phi(z) = \phi_a + \frac{qV}{2} + \frac{z}{d}(\phi_b - qV - \phi_a)$, where $\phi_{a,b}$ are the equilibrium potentials in the absence of the field, and may be taken to be equal to $\phi_a = \phi_b = E_C - E_F$. In this form the left end of the barrier is raised up by $qV/2$, while the right end is lowered by the same amount, see figure 1(c). The tunnelling factor takes the Boltzmann form $e^{-\beta(\frac{\phi_a+\phi_b}{2})}e^{\beta E_z}, E_z = \frac{qV}{2} - \frac{z}{d}qV$. Ignoring unity in the numerator and denominator of the last factor of equation (3) and integrating, we obtain

$$j = -\frac{qm}{\pi^2\hbar^3}k_B T \cdot qVe^{-\beta(E_C-E_T)}\sinh\frac{\beta qV}{2} \qquad (4)$$

There is no current when $V = 0$. Despite the presence of factors of $k_B T$, Eq. (4) does not take into account the radiation loss expressed by the Stefan-Boltzmann law. To incorporate this important feature we must include at the right-hand side of equation (4) the factor $(T_{\text{amb}}/T)^4$, where $T_{\text{amb}}$ is the room temperature.

Equation (4) is incomplete without an independent equation for the temperature. We assume that the temperature $T$ of the electrically active region of the device is spatially uniform. Furthermore, we assume that this temperature is given by [23]

$$T = T_{\text{amb}} + R_{TH}IV, \qquad (5)$$

reminiscent of Newton's law of cooling with $R_{TH}$ [units: $KW^{-1}$] representing the thermal impedance between the electrically active region and the environment. Equations (4) and (5) are now used to plot the current-voltage curves of devices.

## 4. *IV* curves

For our first *IV* plot we take data from M. Le Gallo et al [10] for small doped $Ge_2Sb_2Te_5$ devices but large electric field. The thickness of their amorphous layer was $u_a$ =10.94 nm, the radius of the electrode $r_{\text{BE}}$ =32 nm and the Coulomb centres were $s = 2.6$ nm apart. Because their range of voltage was 2V and the current maximum was only 40$\mu$A, the temperature of the device temperature was effectively constant throughout the measurement. Then the only relevant quantity in Eq. (4) is the barrier height $\phi$. The quantity $V$ in that equation should include the factor $s/u_a$ and, because $j$ is the current density, it must be multiplied by the area $\pi r_{\text{BE}}^2$ to yield current. A thermal impedance of $R_{TH} = 3.7\text{K}\mu\text{W}^{-1}$was assumed. The results are shown in figure 2 and it is clear that the fit is very good. We see negative differential resistance (NDR) on the right-most arm of figure 2 (a). We will discuss this later below.

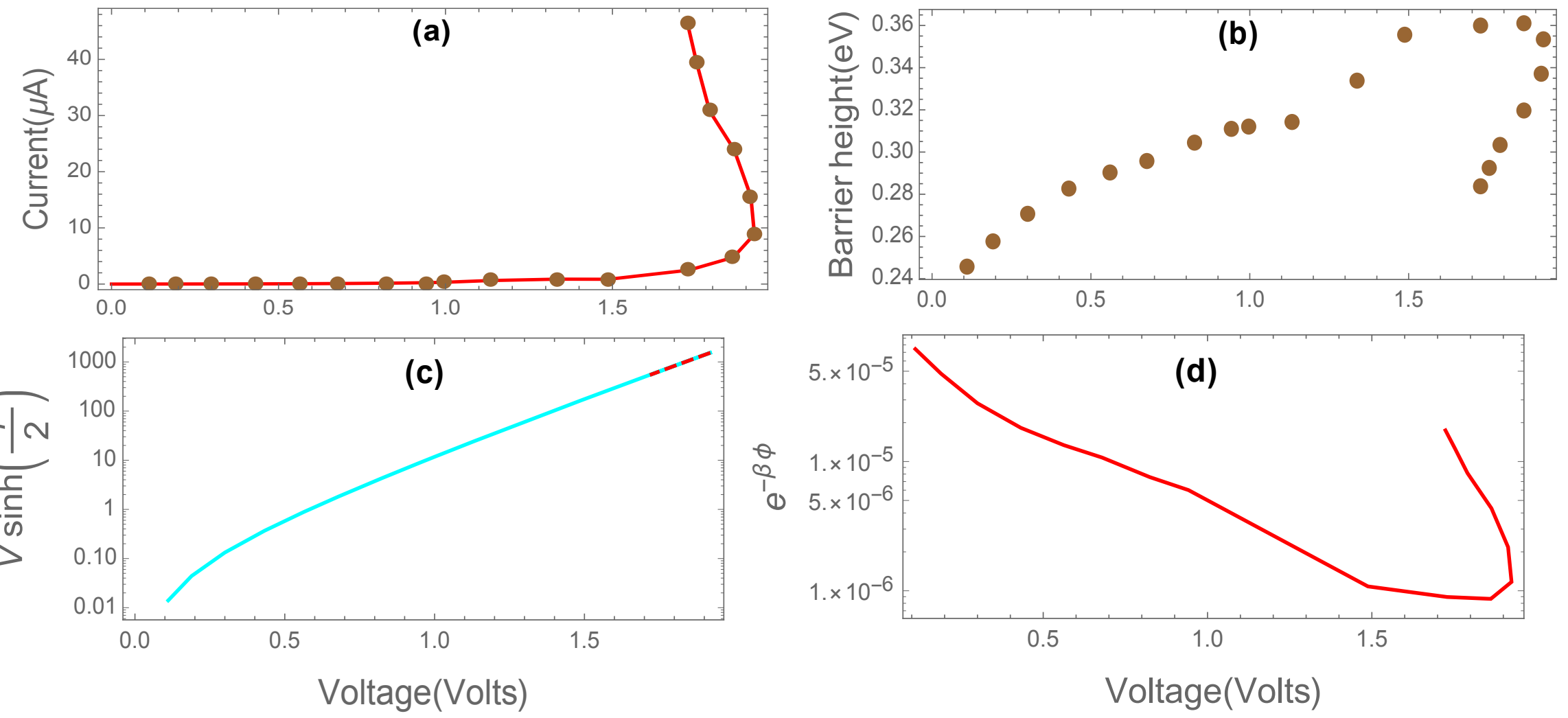

Fig. 2 (a) IV plot based on equation (4) at room temperature. Data points are from ref. 10. (b) Barrier heights used in the fits. (c) & (d) Plots of factors entering into equation (4). Note that temperature varies by less than 1K. The dashed brown line in (c) is the portion of the IV corresponding to NDR and goes backward, retracing the same path as the cyan curve.

We consider next a very large hypothetical device, with an inter-electrode separation of $u_a = 500$ nm. Such large devices are called line-cells and are not only technologically important for memory but also for electrophotonic systems [24, 25]. Figure 3(a) shows the *IV* curve. It is similar to those found by Gwin [26] but the lack of experimental parameters prevented us from further analysis of his results. Negative differential resistance (NDR) is seen when the graph turns backwards. Figure 3(b) displays the temperature of the device as a function of voltage (following equation (5)), while figure 3(c) gives values of the thermal resistance used to obtain figure 3(a). For barrier height $E_C - E_F$ in equation (4) we used values in the limited range 0.266 – 0.271 eV and an inter-trap separation of *s* = 3 nm and a layer cross section area of $4 \times 10^8$ nm$^2$ were assumed. Unlike the previous case where the amorphous element is just 11nm thick, the potential barrier height for this device is relatively insensitive. The starting temperature was taken as $T_{amb}$ = 300 K and the highest temperature attained was ~ 420 K which is well below the critical temperature of ~ 500 K for GeTe [15]. Ideally the thermal resistance $R_{TH}$is constant during the measurement but we note a 20% variation about the average of $150 \text{KW}^{-1}$.

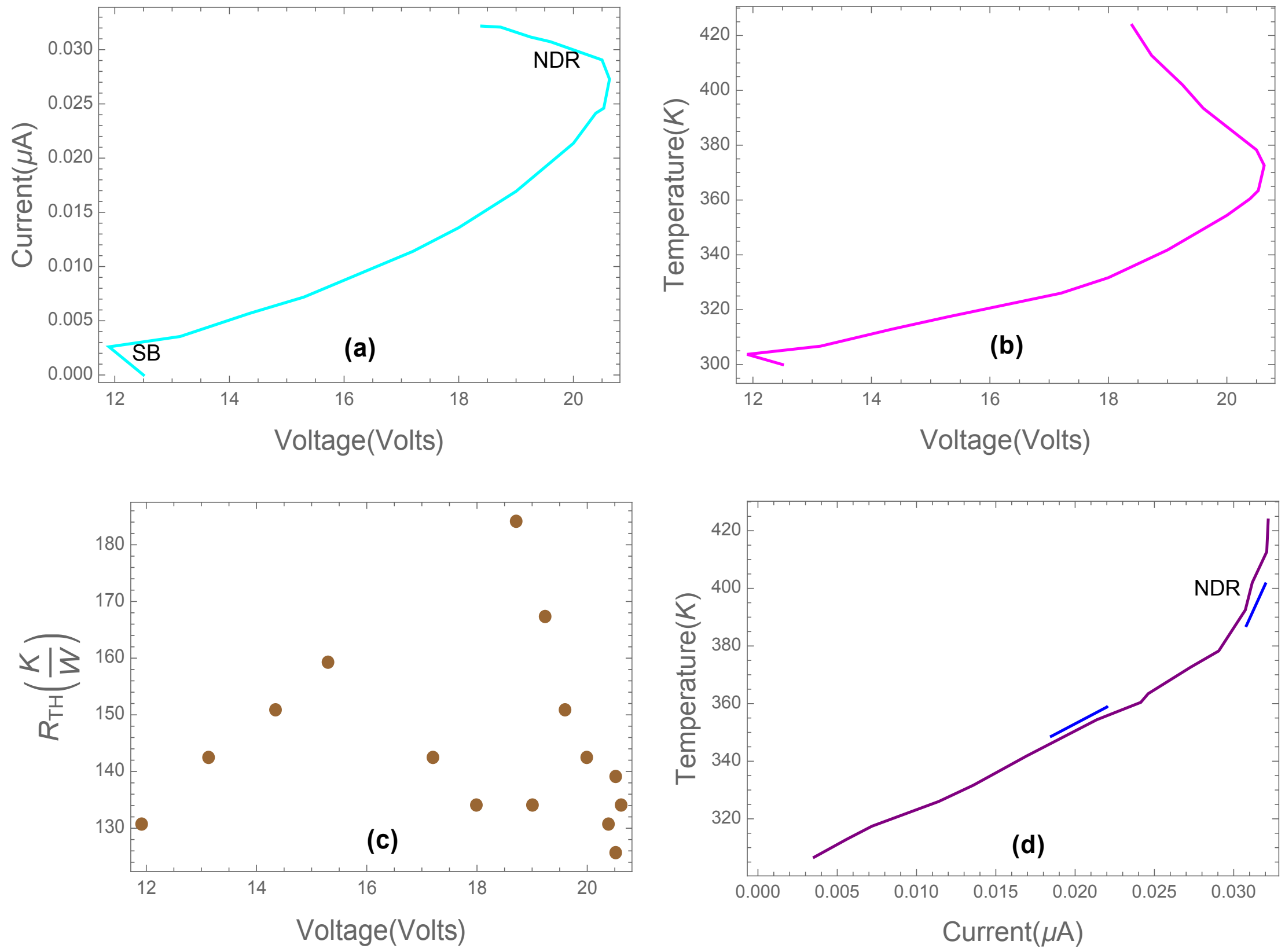

Figure 3 (a) *IV* plot for *a*-GeTe device using equations (4) and (5); (b) corresponding device temperature *T* – *V* plot; (c) values of the thermal impedance $R_{TH}$ for corresponding voltages used in equation (5). (d) *IT* plot shows the device heating up over 4 times faster (as suggested by the blue lines) in the NDR region than in the region before it.

The snapback (SB) is seen at the very beginning of the measurements. We can understand it by considering figures 3(a) and (b) together. Before the onset of conduction, a voltage of 12.5 V is applied to the device. However current and therefore Joule heating is negligible. Therefore the sample has effectively the same temperature as the environment, $T_{\text{amb}} = 300$K. As soon as the current flows the voltage snaps back. In figure 3(a) the voltage dropped to 11.9 V with an accompanying current of ~2.6 mA flowing; and from figure 3(b), we see the temperature concomitantly increasing by about $\Delta T$~3 K. Consider now the exponential factor of equation (4), which is effectively, $e^{-\beta(E_C-E_T)+\frac{\beta qV}{2}}$, since the term $e^{-\frac{\beta qV}{2}}$ of the sinh function is small. During snapback, the device temperature $T$ increases by $\Delta T$, while the potential $V$ dropped by $\Delta V \approx 0.6$V. Thus the exponential changed during snapback to

$$e^{-\beta\left(1-\frac{\Delta T}{T}\right)(E_C-E_T)+\beta\left(1-\frac{\Delta T}{T}\right)\frac{q(V-\Delta V)}{2}} = \left\{e^{-\beta(E_C-E_T)+\frac{1}{2}\beta qV}\right\} e^{\beta\left(\frac{\Delta T}{T}\right)\left(E_C-E_T-\frac{qV}{2}\right)-\frac{1}{2}\beta q\Delta V}. \qquad (6)$$

The second exponential on the right, which represents the change due to the snapback, would have *no* effect if its exponent vanished. That is, the device would see no difference between the situation at the onset of current flow and at the snapback. From this criterion, we can calculate the potential drop $\Delta V$

$$q\Delta V \approx 2\left(\frac{\Delta T}{T}\right)\left(E_C - E_T - \frac{qV}{2}\right) > 0 \qquad (7)$$

With $E_C - E_T \approx 0.27$eV, $qV = 12 \times \frac{3\times10^{-9}}{500\times10^{-9}} = 0.072$eV and $\frac{\Delta T}{T} \approx \frac{3}{300}$, we obtain $q\Delta V \approx 4$ meV, which matches the $12.5 - 11.9 = 0.6$V voltage drop at snapback (equivalent to $\approx 3.6$ meV). Thus the snapback corresponds to the temperature increase of the active layer.

An interest observation is negative differential resistance: it occurs in the *IV* plot (figure 3(a)) where $\frac{dV}{dI} = 0$ and, simultaneously, $\frac{d^2V}{dI^2} < 0$. In figure 3(d) we observe that at the NDR region the device is heats four times faster than prior to it. When high electric fields are applied to the PCM, as in figure 2, there is little temperature change during NDR. This must mean that the barrier height decreases significantly during NDR. In fact figure 2(d) shows that the factor multiplying the barrier grows 10 fold in the NDR region while the remaining factor of equation (4), see the brown line figure 2(c), is only halved in the same interval. In contrast, when low electric fields are applied, as in the case of figure 3, we already know that the barrier height changes only slightly even during NDR whereas the electric current rapidly increases, These observations suggest possible new effects which should be studied in a future work.

## 5. Noise

We finally turn to noise measurements. The power spectral noise due a random signal was given by Machlup [27]

$$S(\omega) = \frac{1}{\pi}\frac{\sigma\tau}{(\sigma+\tau)^2}\frac{\left(\frac{1}{\tau}+\frac{1}{\sigma}\right)}{\omega^2+\left(\frac{1}{\tau}+\frac{1}{\sigma}\right)^2} \qquad (8)$$

where $\sigma$ and $\tau$ are trap emission and capture times and $f = \omega/2\pi$ the frequency. It describes a purely random signal in which emission and capture by traps are uncorrelated. The average power $\Phi(\omega)$ is found by integrating successively over $\sigma$ and $\tau$ through their distribution laws. Assuming a uniform distribution of depths of traps, one finds the distribution laws $p(\sigma) = \frac{1}{\ln\sigma_2/\sigma_1}\frac{1}{\sigma}$ and $p(\tau) = \frac{1}{\ln\tau_2/\tau_1}\frac{1}{\tau}$, where $\sigma_{1,2}$ and $\tau_{1,2}$ are largest and smallest values of these parameters. The result is [28]

$$\Phi(\omega) = \frac{1}{\ln\frac{\sigma_2}{\sigma_1}}\frac{1}{\ln\frac{\tau_2}{\tau_1}}\frac{1}{2\pi\omega^2}\left\{\sum_{i=1,2}\sum_{j=1,2}(-)^j\left(\frac{1}{\sigma_j}+\frac{1}{\tau_i}\right)\log\frac{(\tau_i+\sigma_j)^2}{\sigma_j^2(1+\omega^2\tau_i^2)+2\tau_i\sigma_j+\tau_i^2}\right\}$$
$$+\frac{1}{\ln\frac{\sigma_2}{\sigma_1}}\frac{1}{\ln\frac{\tau_2}{\tau_1}}\frac{1}{\pi\omega}\left\{\tan^{-1}\frac{\omega\sigma_2^2(\tau_2-\tau_1)}{\tau_2\tau_1(1+\omega\sigma_2^2)+\sigma_2\tau_2+\sigma_2\tau_1+\sigma_2^2} - \tan^{-1}\frac{\omega\sigma_1^2(\tau_2-\tau_1)}{\tau_2\tau_1(1+\omega\sigma_1^2)+\sigma_1\tau_1+\sigma_1\tau_2+\sigma_1^2}\right\} \qquad (9)$$

which shows $1/f$ and $1/f^2$ components. The power is symmetric with respect to interchange of $\sigma_{1,2}$ and $\tau_{1,2}$.

The intermittency of resistance noise is essentially a property of the randomness in the appearance and disappearance of electrons or holes. In the case of amorphous materials there is no long-range order, only short-range order. For this reason, the Bloch theorem is inapplicable [7]. Rather, we can envisage finite trains of charges flowing in the conduction band that 'de-rail' and 'restart.' Since trains of charges involve many carriers, it is natural to attribute very low frequencies to these events and concurrently much larger energies than single carriers. Compared with emission and trapping of *single* carriers these de-railings and restarts for trains of carriers are less correlated and hence would be truly random events. Thus, we can expect $1/f$ noise in the very low frequency region $f \lesssim 1$ Hz of the spectrum. A recent measurement reveals a $\frac{1}{f^{\gamma}}$, with $\gamma = 0.9$ noise spectrum for $Ge_2Sb_2Te_5$ [12]. Figure 4 displays a comparison of equation (9) with results from *a*-GeTe measurements at room temperature [29].The log terms in equation (9) are effectively negligible so only the arctangent terms are important. We see that the agreement between the data and equation (9) starts to fray beyond frequencies of 100 Hz. Above this threshold, the $\gamma = 0.9$ value for the exponent becomes appropriate as suggested by the orange line in figure 3. This same observation is also seen in more recent data reported in ref. 12. This effect is no longer due to the de-railing and re-starting of *trains* of charges but due to the electrons or holes.

We saw in equation (1) that information about the electric field is contained in the wave function of the moving charge carrier. Thus, a packed collection of carriers moving under the influence of an electric field share a common phase relation. This relation leads to a form of synchrony between the carriers. A simple tractable model is the Kuramoto model [30, 31], which couples phase oscillators through the phase difference between successive carriers. The model's predictions span the spectrum of full, partial and absence of coherence in the collection. Applying this model to our case, we imagine chains of moving charge carriers of random quantities of electrons/holes, each coherently linked with others in the chain, disappearing and reappearing as they are absorbed into and re-emitted from the traps. If we picture the carriers as masses linked by springs, we see that the total energy scales with the number of masses while the frequency of the system is inversely proportional to the number of masses. That is, the product of energy and frequency is constant. This is precisely the $1/f$ noise effect.

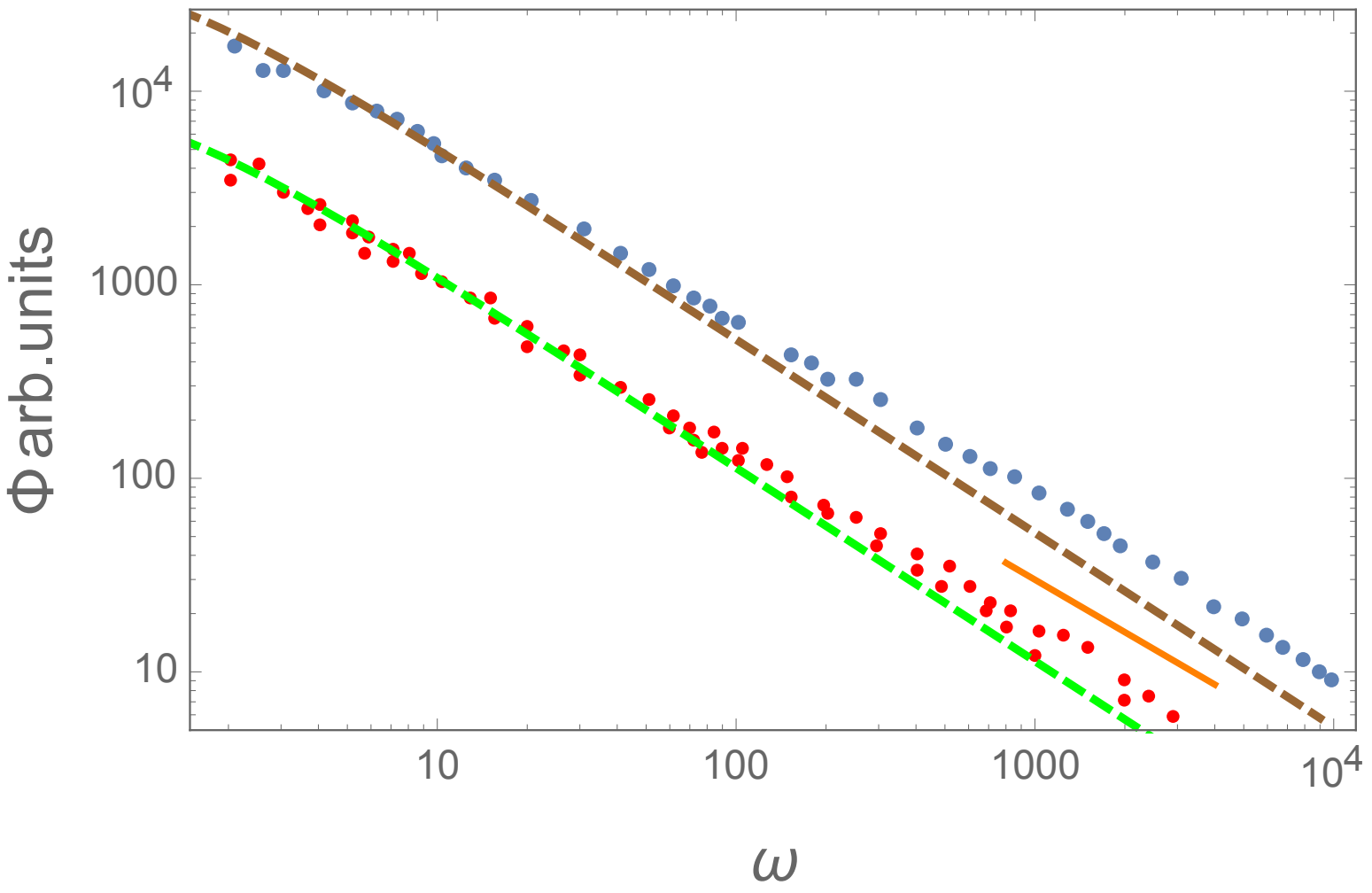

Figure 4 Comparison of data at fixed temperature from ref. 26 with equation (9) (dashed, green and brown lines): parameters are $\sigma_1 = 10^{-6}$ s, $\sigma_2 = 2.5$ s and $\tau_1 = 10^{-6}$ s, $\tau_2 = 2.5$ s. The short orange line at the bottom, right represents a $1/f^{0.9}$ line.

## 6. Conclusions

In this paper, we discussed the role played by temperature in explaining the *IV* plot of amorphous materials. We supplement the PF conduction model with considerations of temperature and provided a reasonable picture of the conduction mechanism in amorphous PCMs. We saw how it could reproduce the *IV* curve for a small $Ge_2Sb_2Te_5$ device with a large field; and how it led to a simple explanation of the snapback in threshold switching for a large device and small field. We demonstrated the versatility of equation (4) over a very large device and studied the region of negative differential resistance. We also discussed how low frequency noise originates from the lack of long-range order in amorphous materials, which in turn causes moving trains of charge carriers to derail and restart *en* route to their end point.

This work was carried out under the auspices of the International Design Center (IDC). The work was sponsored by the Singapore Ministry of Education, grant number: MOE2017-T2-1-161, "Electric-field induced transitions in chalcogenide monolayers and superlattices", and the "Adaptive GHz Devices" project, which is sponsored by Office of Naval Research Global (grant number: N62909-19-1-2005). We are grateful for their financial support and assistance.